\documentclass[12pt]{iopart}
\usepackage{iopams,setstack}
\usepackage{stmaryrd}
\usepackage{amssymb,amsthm}
\usepackage{graphicx,float}
\usepackage[colorlinks=true,linkcolor=blue,citecolor=blue,urlcolor=blue]{hyperref}
\usepackage{color}
\newcommand{\dsp}{\displaystyle}

\newcommand{\Rbb}{\mathbb{R}}
\newcommand{\VecMat}{\ensuremath{\mathrm{Vec}}}
\newcommand{\funf}{\mathcal{F}}

\newcommand{\Cfrak}{\mathfrak{C}}
\newcommand{\Mfrak}{\mathfrak{M}}

\newcommand{\Wfrak}{\mathfrak{W}}

\newcommand{\ep}{e_p}

\newcommand{\diffC}{\dot C}
\newcommand{\tC}{\tilde{C}}
\newcommand{\tCmeas}{\tilde{C}^{\mathrm{meas}}}
\newcommand{\Kexact}{K^{\mathrm{exact}}}

\newcommand{\intint}[2]{\ensuremath{\llbracket#1,#2\rrbracket}}
\newcommand{\intintonen}{\intint{1}{n}}
\newcommand{\intintonentwo}{\intint{1}{n^2}}

\newcommand{\ppmod}[1]{{\!\!\!\!\pmod{#1}}}

\makeatletter
\newcommand{\bigintss}{\@ifnextchar_\@bigintsssub\@bigintssnosub}
\def\@bigintsssub_#1{\def\@int@subscript{#1}\@ifnextchar^\@bigintsssubsup\@bigintsssubnosup}
\def\@bigintsssubsup^#1{\mathop{\text{\LARGE$\int_{\text{\normalsize$\scriptstyle\kern-0.25em\@int@subscript$}}^{\text{\normalsize$\scriptstyle#1$}}$}}\nolimits}
\def\@bigintsssubnosup{\mathop{\text{\LARGE$\int_{\text{\normalsize$\scriptstyle\@int@subscript$}}$}}\nolimits}
\def\@bigintssnosub{\@ifnextchar^\@bigintssnosubsup\@bigintssnosubnosup}
\def\@bigintssnosubsup^#1{\mathop{\text{\LARGE$\int^{\text{\normalsize$\scriptstyle#1$}}$}}\nolimits}
\def\@bigintssnosubnosup{\mathop{\text{\LARGE$\int$}}\nolimits}
\makeatother

\newcommand{\be}{\begin{equation}}
\newcommand{\ee}{\end{equation}}

\newcommand{\bes}{\begin{equation*}}
\newcommand{\ees}{\end{equation*}}

\begin{document}
\title[Compartmental analysis]{Compartmental analysis of dynamic nuclear medicine data: regularization procedure and application to physiology}
\author{Fabrice Delbary$^1$ and Sara Garbarino$^2$}

\address{$^1$ Institut f\"{u}r Mathematik, Johannes Gutenberg-Universit\"at Mainz, Mainz \\
$^2$ Centre for Medical Image Computing, Department of Computer Science, University College London, London}

\begin{abstract}
Compartmental models based on tracer mass balance are extensively used
in clinical and pre-clinical nuclear medicine in order to obtain quantitative information
on tracer metabolism in the biological tissue. This paper is the second of a series of two that deal with the problem of tracer coefficient estimation via compartmental modelling
in an inverse problem framework. While the previous work was devoted to the discussion of identifiability issues for $2$, $3$ and $n$-dimension compartmental systems, here we discuss the problem of numerically determining the tracer coefficients by means of a general regularized Multivariate Gauss Newton scheme. In this paper, applications concerning cerebral, hepatic and renal functions are considered, involving experimental measurements on FDG--PET data on different set of murine models.
\end{abstract}

\section{Introduction}

Nuclear medicine imaging is a class of functional imaging modality that utilizes radioactive tracers to investigate specific physiological processes. Such tracers are in general short-lived isotopes that are injected in the subject's blood and linked to chemical compounds whose metabolism is highly significant to understand the function or malfunction of an organ. Positron Emission Tomography (PET) \cite{ollinger1997positron} is the most modern nuclear medicine technique, utilizing isotopes produced in a cyclotron and providing dynamical images of its metabolism-based accumulation in the tissues. 

Applications of PET in the clinical workflow depend on the kind of tracer employed and on the kind of metabolism that such tracer is able to involve: in this paper we will make use of $[^{18}$ F]fluoro-2-deoxy-D-glucose--PET (FDG--PET) data. FDG is largely used for PET, mainly in the case of oncological applications \cite{antoch2004accuracy, avril2001glucose, ziegler1999reproducibility, delbeke2006procedure}. 

This paper describes a very general numerical scheme for the reduction of different compartment models of FDG metabolism, based on a regularized multivariate Gauss Newton algorithm \cite{engl1996regularization}. As many other approaches to compartmental analysis \cite{garbarino2013estimate,garbarino2014novel,gunn2001positron,kamasak2005direct,qiao2007kidney,schmidt2002kinetic,sourbron2011tracer}, also this method realizes numerical optimization but in a peculiarly effective way: the matrix differentiation step required at some stage of the analysis is here performed analytically, thus avoiding time consuming numerical differentiation. We test the reliability of the approach in the cases of a synthetic dataset and of three sets of experimental measurements concerned with the FDG metabolism in the brain, in the liver, and in the kidneys, provided by a micro--PET scanner for small animals.

The plan of the paper is as follows. Section 2 briefly recall the general $n$--compartment model
for FDG metabolism. Section 3 introduces the numerical method for model reduction. Section
4 describes the FDG--PET models for the cerebral, hepatic and renal physiology. Section 5 provides numerical and experimental validation of the approach. Our conclusions are offered
in Section 6.\\

In the document, $\Rbb_+,\Rbb_-,\Rbb^*_+,\Rbb^*_-$ respectively denote the set of non-negative real numbers, the set of non-positive real numbers, the set of positive real numbers, the set of negative real numbers. For non-negative integers $p$ and $q$, $\intint{p}{q}$ denotes the set of integer larger than or equal to $p$ and smaller than or equal to $q$. For a positive integer $n$, the canonical basis of $\Rbb^n$ is denoted by $(\ep)_{p\in\intintonen}$ and $M_n(\Rbb)$ denotes the algebra of $n\times n$ matrices with coefficients in $\Rbb$. For Banach or Fr\'echet spaces $E$ and $F$, the vector space of bounded operators from $E$ to $F$ is denoted by $L(E,F)$.

\section{The $n$-compartment system}
We introduce the more general case of an $n$--compartment system and then describe the regularized Multivariate Gauss Newton algorithm. Simulation results and real data results will be provided on the more reliable cases of two--compartment and three--compartment models. The identifiability results of \cite{delbary} insures that in the cases we consider, we can have good expectations of robusteness and accuracy even though the initial guess is not well--chosen, as applications can prove. However, in more general cases, either it is known that no uniqueness holds or when no information on uniqueness is known, the robustness and accuracy of the algorithm might greatly rely on the choice of a good initial guess.
\begin{figure}
\begin{center}
\includegraphics[scale=0.625]{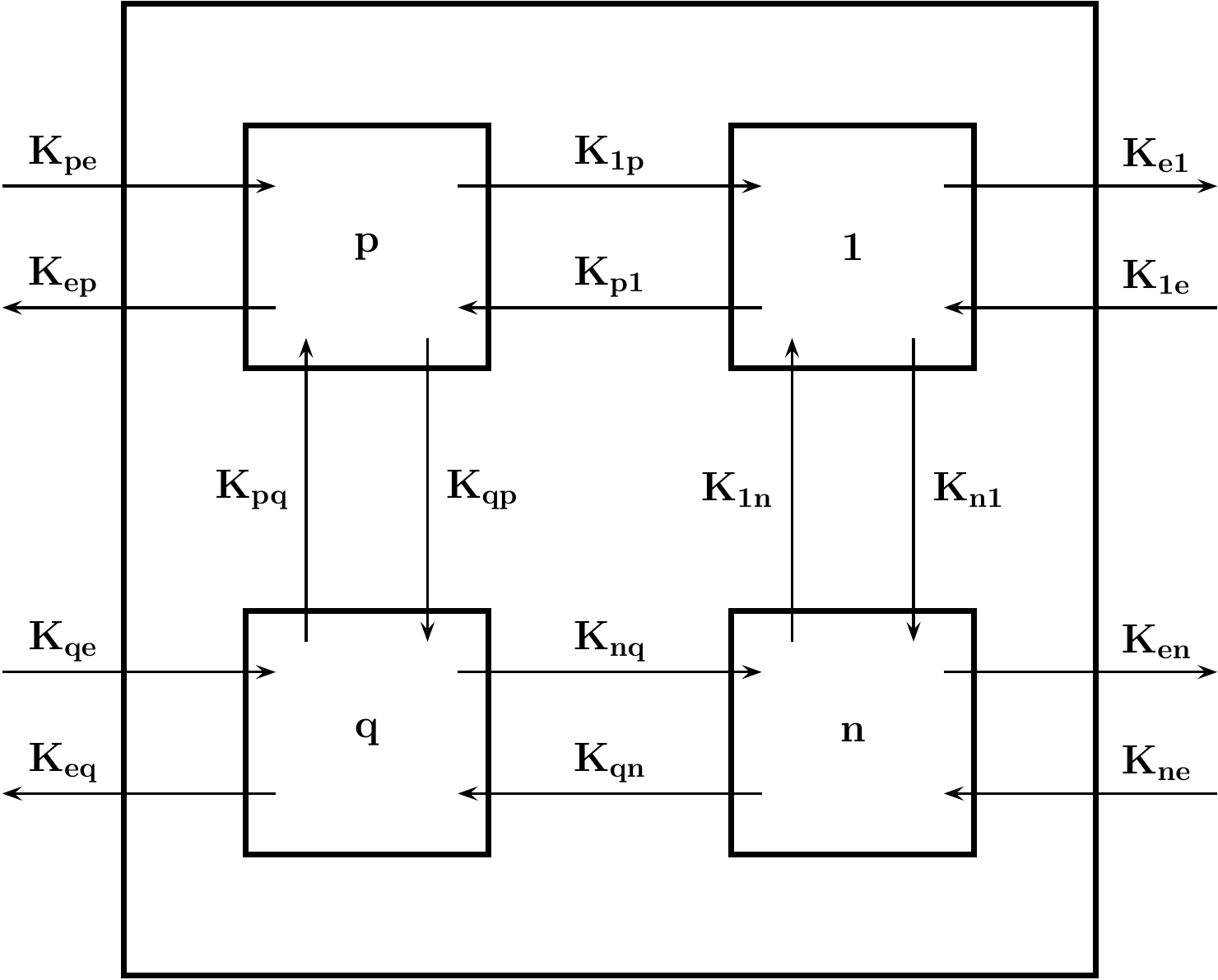}
\caption{$n$-compartment system}\label{n-compartment-model}
\end{center}
\end{figure}

We consider a system composed of $n$ compartments as in Figure \ref{n-compartment-model}, typically an organ, where each compartment represents a functional behaviour of a tissue, or an anatomical district. Remark that although the generic term ``compartment'' is used, it does not necessary mean that each compartment is contained in a physical compartment distinguishable from the others. In fact, this even constitutes one of the main issue in the inverse problem of getting information on the system since each compartment can not be individually observed. A radioactive tracer is injected to a patient and for a compartment $p\in\intintonen$, $C_p$ denotes the non-negative concentration function of the tracer in the compartment. The compartment $p$ receives the radioactive tracer from the outside world at a constant non-negative rate $k_{p\mathrm{e}}$ and a non-negative concentration function $C_{p\mathrm{e}}$ and it excretes the tracer at a constant non-negative rate $k_{\mathrm{e}p}$ in the outside world. The constant non-negative rate at which the compartment $p$ receives the tracer from a compartment $q\neq p$ is 
denoted $k_{pq}$. The concentration functions $(C_{p\mathrm{e}})_{p\in\intintonen}$ are supposed to be continuous. The evolution of the tracer concentrations in each compartment is then governed by the following linear system of ordinary differential equations with constant coefficients
\bes
\diffC_p=\sum_{q=1}^n k_{pq}C_q+k_{p\mathrm{e}}C_{p\mathrm{e}},\qquad p\in\intintonen,
\ees
with the initial conditions
\be\label{eq:fwdpb}
C_p(0)=0,\qquad p\in\intintonen,
\ee
where, for $p\in\intintonen$, $k_{pp}=-\left(\sum_{q\neq p}k_{qp}+k_{\mathrm{e}p}\right)$. That is
\bes
\diffC=MC+W,\qquad C(0)=0, 
\ees
where
\be\label{vecCvecrhs}
C=\pmatrix{C_1\cr
\vdots\cr
C_n\cr},
\qquad
W=\pmatrix{k_{1\mathrm{e}}C_{1\mathrm{e}}\cr
\vdots\cr
k_{n\mathrm{e}}C_{n\mathrm{e}}\cr},
\ee
and the matrix $M$ is given by
\be\label{sysmat}
M_{pq}=k_{pq},\qquad p,q\in\intintonen.
\ee
PET-scan images allow to know the total amount of radioactive tracer in the organ. Each compartment contributes to the intensity of the image linearly with respect to the amount of tracer in the compartment, that is its volume times the tracer concentration. Hence, in practice, PET-scan data gives access to
\be\label{eq:data}
\tC(t)=\alpha^TC(t),\qquad t\in\Rbb_+,
\ee
where $\alpha\in{\Rbb_+^*}^n$ is a known constant vector. Thus, the inverse problem we consider is to recover the exchange rates $\underline{K}\in\Rbb^{n^2+n}$, where
\be\label{defveck}
\underline{K}_p=\cases{k_{\mathrm{e}p},&$p\in\intintonentwo,\;p\equiv1\ppmod{n+1}$,\\
k_{p-n\left\lfloor\frac{p-1}{n}\right\rfloor,1+\left\lfloor\frac{p-1}{n}\right\rfloor},&$p\in\intintonentwo,\;p\not\equiv1\ppmod{n+1}$,\\
k_{(p-n^2)\mathrm{e}}&$p\in\intint{n^2+1}{n^2+n}$,}
\ee
with measures of $\tC$. Note however that some exchange rates may be \textit{a priori} known.

\section{A Newton algorithm for the inverse problem}

In the following of the document, for a positive integer $n$ 
and $K\in\Rbb^{n^2+n}$, we denote by $\hat{K}\in\Rbb^{n^2}$ the first $n^2$ components of $K$ and $\check{K}\in\Rbb^n$ the last $n$ components of $K$. For a positive integer $n$, we denote by $\Mfrak$ the following linear operator
\bes
\begin{array}{rcl}
\Mfrak:\Rbb^{n^2}&\to&M_n(\Rbb)\\\
H&\mapsto&\Mfrak(H),
\end{array}
\ees
where for all $H\in\Rbb^{n^2}$
\bes
\fl\Mfrak(H)_{pq}=\cases{-H_{1+(n+1)(p-1)}-\sum_{\scriptsize\begin{array}{l}p'=1\\p'\neq p\end{array}}^nH_{p+n(p'-1)},&$p,q\in\intintonen,\; p=q$,\\
H_{p+n(q-1)},&$p,q\in\intintonen,\;p\neq q$,}
\ees
so that for all $H\in\Rbb_+^{n^2}$, $\Mfrak(H)$ is the matrix defined in \eref{sysmat} for the parameters $H$. Consider now the linear operator 
$\VecMat:M_n(\Rbb)\to\Rbb^{n^2}$ stacking the columns of a matrix $A$ into a 
column vector $\VecMat(A)$, that is
\be\label{defvecmat}
\VecMat(A)_p=A_{p-n\left\lfloor\frac{p-1}{n}\right\rfloor,1+\left\lfloor\frac{
p-1}{n}\right\rfloor},\qquad p\in\intintonentwo.
\ee
The operator $\VecMat\circ\Mfrak$ is linear from $\Rbb^{n^2}$ into itself, we 
denote by $S\in M_{n^2}(\Rbb)$ its matrix. It is a $n^2\times n^2$ sparse matrix 
with $2n^2-n$ non-zero elements which are equal to $\pm1$. More precisely, 
$n^2-n$ entries of $S$ are equal to $1$ and $n^2$ entries of $S$ are equal to 
$-1$, such as
\be\label{matrixs}
S_{pq}=\cases{1,&$p,q\in\intintonentwo,p=q,p\not\equiv1\ppmod{n+1}$,\\
-1,&$p,q\in\intintonentwo,p\equiv1\ppmod{n+1},p\equiv q\ppmod{n}$,\\
0,&otherwise.}
\ee
We denote by $\Wfrak$ the following linear operator
\bes
\begin{array}{rcl}
\Wfrak:C^0(\Rbb_+,\Rbb)^n&\to&L(\Rbb^n,C^0(\Rbb_+,\Rbb)^n)\\
\check{C}&\mapsto&\Wfrak(\check{C}),
\end{array}
\ees
where for all $\check{C}\in C^0(\Rbb_+,\Rbb)^n$, $H\in\Rbb^n$ and $p\in\intintonen$, $[\Wfrak(\check{C})(H)]_p=H_p\check{C}_p$, so that for all vector of input concentrations functions $\check{C}=(C_{p\mathrm{e}})_{p\in\intintonen}\in C^0(\Rbb_+,\Rbb_+)^n$ and $H\in\Rbb_+^n$, $\Wfrak(\check{C})(H)$ is the vector defined in \eref{vecCvecrhs} for the input concentrations functions $\check{C}$ and the parameters $H$. We denote by $\Cfrak$ the following function
\bes
\begin{array}{rcl}
\Cfrak:C^0(\Rbb_+,\Rbb)^n&\to&L(\Rbb^{n^2+n},C^1(\Rbb_+,\Rbb)^n)\\
\check{C}&\mapsto&\Cfrak(\check{C}),
\end{array}
\ees
where for all $\check{C}\in C^0(\Rbb_+,\Rbb)^n$ and $K\in\Rbb^{n^2+n}$, $C=\Cfrak(\check{C})(K)\in C^1(\Rbb_+,\Rbb)^n$ is the unique solution to
\bes
\diffC=MC+W,\qquad C(0)=0,
\ees
where $M=\Mfrak(\hat{K})$ and $W=\Wfrak(\check{C})(\check{K})$. For $\alpha\in{\Rbb^*_+}^n$, $\Cfrak^\alpha$ is the function defined by
\bes
\begin{array}{rcl}
\Cfrak^\alpha:C^0(\Rbb_+,\Rbb)^n&\to&L(\Rbb^{n^2+n},C^1(\Rbb_+,\Rbb))\\
\check{C}&\mapsto&\left[K\mapsto\alpha^T(\Cfrak(\check{C})(K))\right].
\end{array}
\ees
\subsection{Algorithm description}
We consider a $n$-compartment system as in Figure \ref{n-compartment-model} 
with known input concentration functions 
$\check{C}=(C_{p\mathrm{e}})_{p\in\intintonen}\in C^0(\Rbb_+,\Rbb_+)^n$ and 
$\alpha\in{\Rbb^*_+}^n$. We recall that the problem we are interested in 
consists in recovering the exchange rates $K\in\Omega\subset\Rbb^{n^2+n}$, where 
$K$ is given by \eref{defveck}, from the knowledge of $\tC(K)$, where 
$\tC=\Cfrak^\alpha(\check{C})$. We recall that
\bes
\tC(K)(t)=\alpha^T C(K)(t),\qquad t\in\Rbb_+,
\ees
where $C=\Cfrak(\check{C})$. The concentrations vector $C(K)$ is the solution 
of the ordinary differential equations
\bes
\diffC(K)=\Mfrak(\hat{K})C(K)+W(\check{K}),\qquad C(K)(0)=0, 
\ees
where $W=\Wfrak(\check{C})$. The solution $C(K$) to the system of ordinary 
differential equations is given by
\bes
C(K)(t)=\int_0^t \exp((t-\tau)\Mfrak(\hat{K}))W(\check{K})(\tau)\,d\tau,\qquad 
t\in\Rbb_+,
\ees
hence
\bes
\tC(K)(t)=\alpha^T\int_0^t\exp((t-\tau)M(\hat{K}))W(\check{K})(\tau)\,d\tau,\qquad t\in\Rbb_+.
\ees
$\tC:\Rbb^{n^2+n}\to C^1(\Rbb_+,\Rbb)^n$ can be easily seen to be differentiable 
and even analytic. In order to use a Newton algorithm, we need to compute its 
Fr\'echet derivative. More precisely, considering $t\in\Rbb_+$, we will compute 
the gradient of $\tC_t$ for all $t\in\Rbb_+$, with respect to $K$, where for all 
$K\in\Rbb^{n^2+n}$, $\tC_t(K)=\tC(K)(t)$. For all $K\in\Rbb^{n^2+n}$, the 
Fr\'echet derivative $\frac{d\tC}{dK}(K)$ of $\tC$ at $K$, bounded operator from 
$\Rbb^{n^2+n}$ to $C^1(\Rbb_+,\Rbb)$ is then given by
\bes
\begin{array}{rcl}
\dsp \frac{d\tC}{dK}(K) : \Rbb^{n^2+n} & \to & C^1(\Rbb_+,\Rbb),\\
H & \mapsto & \left[t \mapsto \nabla \tC_t(K) \cdot H\right].
\end{array}
\ees
For all $t\in\Rbb_+$, the gradient of $\tC_t$ is given by
\be\label{fullgrad}
\nabla\tC_t=
\pmatrix{\nabla_{\hat{K}}\tC_t\cr
\nabla_{\check{K}}\tC_t\cr},
\ee
where $\nabla_{\hat{K}}$ denotes the gradient with respect to $\hat{K}$ and 
$\nabla_{\check{K}}$ denotes the gradient with respect to $\check{K}$. Since 
$\tC_t$ is linear with respect to $\check{K}$, we simply have for all 
$K\in\Rbb^{n^2+n}$
\be\label{gradkcheck}
\nabla_{\check{K}}\tC_t(K)=\left(\int_0^t\check{C}(\tau)\odot\exp((t-\tau)\Mfrak(\hat{K})^T)\,d\tau\right)\alpha,
\ee
where $\odot$ denotes the Khatri-Rao product. Compute now the gradient of 
$\tC_t$ with respect to $\hat{K}$. Writing $\tC_t$ as
\bes
\tC_t(K)=\alpha^T \funf_t(\check{K},\Mfrak(\hat{K})),\qquad K\in\Rbb^{n^2+n},
\ees
where for all $K\in\Rbb^{n^2+n}$ and $N\in M_n(\Rbb)$
\bes
\funf_t(\check{K},N)=\int_0^t \exp((t-s)N)W(\check{K})(s)\,ds,
\ees
we have for all $K\in\Omega$ and $N,H\in\Rbb^{n^2}$
\bes
\nabla_{\hat{K}}\tC_t(K)\cdot H=\alpha^T\frac{\partial\funf_t}{\partial N}\left(\left(\check{K},\Mfrak(\hat{K})\right);\frac{d\Mfrak}{d\hat{K}}(\hat{K};H)\right),
\ees
and since $\Mfrak:\Rbb^{n^2}\to M_n(\Rbb)$ is linear
\be\label{gradkhatbeg}
\nabla_{\hat{K}}\tC_t(K)\cdot H=\alpha^T\frac{\partial\funf_t}{\partial N}\left(\left(\check{K},\Mfrak(\hat{K})\right);\Mfrak(H)\right).
\ee
Hence, to compute the gradient of $\tC_t$ with respect to $\hat{K}$, we need to 
compute the Fr\'echet derivative of $\funf_t$ with respect to the second 
variable $N\in M_n(\Rbb)$. The Fr\'echet derivative of the exponential function 
can be written in this way
\bes
\frac{d\exp}{dN}(N;H)=\int_0^1 \exp(\tau N)H\exp((1-\tau)N)\,d\tau,\qquad 
N,H\in M_n(\Rbb).
\ees
Hence, for all $K\in\Omega$ and $N,H\in M_n(\Rbb)$
\bes
\fl\frac{\partial\funf_t}{\partial N}\left(\left(\check{K},N\right);H\right)=\int_0^t(t-s)\left(\int_0^1\exp(\tau(t-s)N)H\exp((1-\tau)(t-s)N)\,d\tau\right)W(\check{K})(s)\,ds,
\ees
that is, with the change of variables $\tau(t-s)\to\tau$
\bes
\fl\frac{\partial\funf_t}{\partial N}\left(\left(\check{K},N\right);H\right)=\int_0^t\left(\int_0^{t-s}\exp(\tau N)H\exp((t-s-\tau)N)\,d\tau\right)W(\check{K})(s)\,ds.
\ees
Hence
\bes
\fl\frac{\partial\funf_t}{\partial N}\left(\left(\check{K},N\right);H\right)=\int_0^t\left(\int_0^{t-\tau}\exp(\tau N)H\exp((t-s-\tau)N)W(\check{K})(s)\,ds\right)\,d\tau,
\ees
so that making the change of variables $(t-\tau)\to\tau$, we get
\bes
\fl\frac{\partial\funf_t}{\partial N}\left(\left(\check{K},N\right);H\right)=\int_0^t\exp((t-\tau)N)H\left(\int_0^\tau\exp((\tau-s)N)W(\check{K})(s)\,ds\right)\,d\tau.
\ees
In other words
\bes
\frac{\partial\funf_t}{\partial N}\left(\left(\check{K},N\right);H\right)=\int_0^t\exp((t-\tau)N)H\funf_\tau(N)\,d\tau.
\ees
Writing the previous formula in a more convenient way for computations, we have
\bes
\frac{\partial\funf_t}{\partial N}\left(\left(\check{K},N\right);H\right)=\left(\int_0^t\funf_\tau(N)^T\otimes\exp((t-\tau)N)\,d\tau\right)\VecMat(H),
\ees
where $\otimes$ denotes the Kronecker product and the linear operator $\VecMat:M_n(\Rbb)\to\Rbb^{n^2}$, stacking the column of a matrix into a column vector, is defined in \eref{defvecmat}. Consequently, from \eref{gradkhatbeg}, we have for all $K\in\Rbb^{n^2+n}$ and $H\in\Rbb^{n^2}$
\bes
\fl\nabla_{\hat{K}}\tC_t(K)\cdot H=\alpha^T\left(\int_0^t C(K)(\tau)^T\otimes\exp((t-\tau)\Mfrak(\hat{K}))\,d\tau\right)\VecMat(\Mfrak(H)).
\ees
Recalling that $S$, defined in \eref{matrixs}, denotes the matrix of the linear operator $\VecMat\circ \Mfrak:\Rbb^{n^2}\to\Rbb^{n^2}$, we then have
\bes
\nabla_{\hat{K}}\tC_t(K)\cdot H=\alpha^T\left(\int_0^t C(K)(\tau)^T\otimes\exp((t-\tau)\Mfrak(\hat{K}))\,d\tau\right)SH,
\ees
so that
\be\label{gradkhatend}
\nabla_{\hat{K}}\tC_t(K)=S^T\left(\int_0^t C(K)(\tau)\otimes\exp((t-\tau)\Mfrak(\hat{K})^T)\,d\tau\right)\alpha.
\ee
Hence, using \eref{fullgrad}, \eref{gradkcheck}, \eref{gradkhatend}
\bes
\nabla\tC_t(K)=
\bigintss_0^t
\pmatrix{S^T\left(C(K)(\tau)\otimes\exp((t-\tau)\Mfrak(\hat{K})^T)\right)\alpha\cr
\left(\check{C}(\tau)\odot\exp((t-\tau)\Mfrak(\hat{K})^T)\right)\alpha\cr}\,d\tau.
\ees
In other words, for all $K,H\in\Rbb^{n^2+n}$ and $t\in\Rbb_+$, we have
\bes
\fl\left[\frac{d\tC}{dK}(K;H)\right](t)=\left(\bigintss_0^t\pmatrix{S^T\left(C(K)(\tau)\otimes\exp((t-\tau)\Mfrak(\hat{K})^T)\right)\alpha\cr
\left(\check{C}(\tau)\odot\exp((t-\tau)\Mfrak(\hat{K})^T)\right)\alpha\cr}\,d\tau\right)H.
\ees
Consider a known function $\tCmeas\in C^1(\Rbb_+,\Rbb)$, the measures of 
$\tC(\Kexact)$ where $\Kexact$ are the real unknown exchange rates of the 
compartmental system. Let $K^0\in\Omega$ be an initial guess, then the Newton 
algorithm consists in solving the linear equation with unknown $H^0\in\Omega$
\be\label{eq:gennew}
\left[\frac{d\tC}{dK}(K^0;H^0)\right](t)=\tCmeas(t)-\tC(K^0)(t),\qquad\mbox{for all }t\in\Rbb_+.
\ee
Then, increment the value $K^0$, giving $K^1=K^0+H^0$ and iterate the process.
\subsection{Implementation details}
The equation \eref{eq:gennew} may have no solution, moreover, in 
real applications, one has only the measured data $\tCmeas$ for a finite number 
of sampling time points $t_1,\ldots,t_m\in\Rbb_+$ and the data may be noisy. The 
discretized Newton algorithm consists in solving the linear system 
\eref{eq:gennew} by Tikhonov regularization. The non-regularized discretized 
system is given by
\bes
\pmatrix{\left[\nabla\tC_{t_1}(K^0)\right]^T\cr
\vdots\cr
\left[\nabla\tC_{t_m}(K^0)\right]^T\cr}H^0=\pmatrix{
\tCmeas(t_1)-\tC_{t_1}(K^0)\cr
\vdots\cr
\tCmeas(t_m)-\tC_{t_m}(K^0)\cr},
\ees
that is, denoting by $A_0$ the matrix
\bes
A_0=
\pmatrix{\left[\nabla\tC_{t_1}(K^0)\right]^T\cr
\vdots\cr
\left[\nabla\tC_{t_m}(K^0)\right]^T\cr},
\ees
and by $Y^0$ the vector
\bes
Y^0=\pmatrix{\tCmeas(t_1)-\tC_{t_1}(K^0)\cr
\vdots\cr
\tCmeas(t_m)-\tC_{t_m}(K^0)\cr},
\ees
the non-regularized discretized system can be written as
\be\label{eq:gennewdisc}
A_0H^0=Y^0.
\ee
Solving the system \eref{eq:gennewdisc} by Tikhonov regularization consists in 
finding the solution $H^0$ to
\bes
(rI+A_0^TA_0)H^0=A_0^TY^0,
\ees
where $r$ is a regularization parameter. As previously described, the value 
$K^0$ is incremented, giving $K^1=K^0+H^0$ and the process is iterated. Note 
that the regularization parameter can be different at each iteration step.

\section{The FDG--PET models}

\begin{figure}[H]
\begin{center}
\includegraphics[scale=0.75]{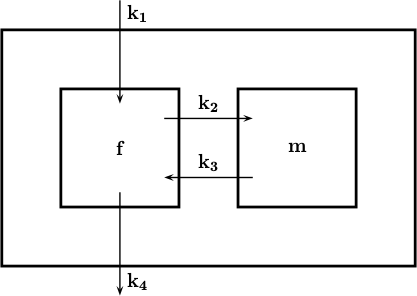}
\caption{$2$-compartment catenary model}\label{fig:2cmpgen}
\end{center}

\end{figure}
As already mentioned, FDG is largely used for PET, mainly in the case of oncological applications. In fact, FDG-PET is based on the higher glycolytic activity in tumor cells compared to healthy tissue as the cause of image contrast. This glucose--analog behaviour implies that FDG is transported into malignant cells, which therefore exhibit increased radioactivity. From a biochemical viewpoint, FDG may follow a two-destiny path: in the first path FDG molecules are phosphorylated by means of a 6-phosphate group and remain trapped within the cells while, in the second path, FDG is not metabolized and therefore remains free in the tissue. This behaviour implies that, from a compartmental perspective, the analysis of FDG-PET data relies on a model made of an Input Function (in general, but not exclusively, an arterial one) and two functional compartments describing the free tracer and the trapped, metabolized FDG-6P, respectively. 

In the two-compartment model of FDG-PET data (see \figurename~\ref{fig:2cmpgen}) the state variables are the tracer concentration in compartments $C_f$ and $C_m$ for the free and metabolized compartments, respectively, while the kinetic process in the system is initialized by the Input Function (IF) $C_b$, representing the tracer concentration in blood. The four constant transmission coefficients between the communicating compartments are denoted, for sake of simplicity, as $k_1,\,k_2,\,k_3, \,k_4$. The model equations for the forward and inverse problem are easily obtained from the general form in eq (\ref{eq:fwdpb})--(\ref{defveck}).

Next subsections describe three compartmental models for the FDG related to different physiological systems.

\subsection{The model of the brain}

For the FDG model of the brain, we will use a two--compartment catenary compartmental model. Indeed, even if from a functional (neuronal) viewpoint the brain has an extremely complex behaviour, brain--glucose metabolism inside the brain can be thought as standard \cite{alf2013quantification}. 
We are aware that recently has been shown that the brain exchange coefficients of the compartmental systems, can vary very much according to the spatial position in the organ. For this reason, the two--compartment catenary compartmental system used for describing the brain physiology is modelled, applied and reduced \emph{pixelwise} \cite{karakatsanis2013dynamic, wardak2013simplified, angelis2014evaluation}. Such compartmental models are known as \emph{parametric compartmental models} or \emph{indirect parametric imaging}.

We will not discuss such parametric models, and otherwise discuss the validity and robustness of our technique if applied to a two--compartment system, using brain data to validate it on real imaging FDG--PET data.

\subsection{The model of the liver}
\begin{figure}[H]
\begin{center}
\includegraphics[scale=0.75]{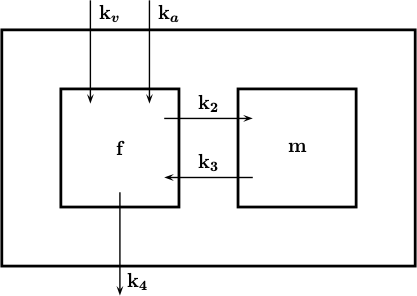}
\caption{$2$-compartment model with $2$ input functions (liver)}\label{fig:liver}
\end{center}
\end{figure}
For liver studies, the FDG--PET compartmental analysis is complex, due to the hepatocyte capability to de--phosphorilate and release glucose and FDG into the blood. Moreover, a tracer is supplied to the liver by both the hepatic artery and the portal vein (dual input). Therefore the compartmental model that we need to adopt, is a two--compartment catenary one, with two IFs. However, the portal vein is not visible in PET images; in order to overcome this limitation we coupled the capability of PET to provide FDG concentration curves in virtually all the organs with the anatomical ken that the portal vein enters the liver just after leaving the gut, as in \cite{garbarino2015new}. Indeed, we use this information to reduce a compartmental scheme in which gut FDG metabolism is modelled as a standard two--compartment compartmental system, its tracer concentration is used to compute the portal vein IF, and therefore a model as in Figure \ref{fig:liver} is reduced, to describe tracer kinetics in the hepatic system.

\subsection{The model of the kidneys}

\begin{figure}[H]
\begin{center}
\includegraphics[scale=0.75]{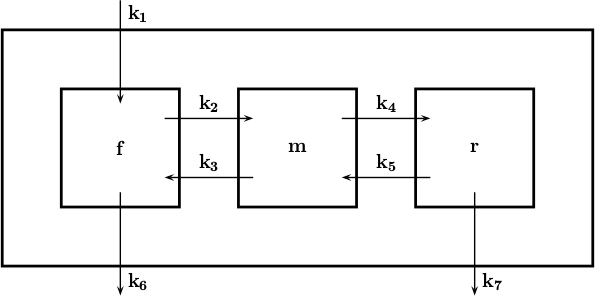}
\caption{$3$-compartment model for kidneys}
\end{center}
\end{figure}

Clinical activity asks to represent glucose consumption by malignant cells as Standardized Uptake Value (SUV) \cite{delbeke1999optimal} whose estimation is hampered by the interference of several variables modulating tracer availability independently of tumor metabolism \cite{diederichs1998fdg}. Among these factors, kidney function plays a relevant role as, differently from the tracked glucose, FDG is scarcely reabsorbed in renal tubule and is largely excreted in the urines \cite{shreve1999pitfalls}. Furthermore, variations in FDG concentration due to water re-absorption in renal tubules and increase of bladders
volume during the FDG excretion process have to be taken in account, in order to describe the renal FDG metabolism: therefore we describe the renal system as a three--compartmental system. According to \cite{garbarino2014novel}, we have some physiological constraints that simplify the compartmental scheme: $k_7$ is proportional to $k_4$ with a known proportional factor $\lambda = 10^2$, i.e. $k_4 = 10^2\ k_7$; $k_5$ is set to zero since the tracer is not reabsorbed in tubules and goes to bladder (which is the output $k_7$ of the compartmental scheme). 
\section{Numerical Results}
An 'Albira' micro--PET system produced by Carestream Health is currently operational at the IRCCS San Martino IST, Genova. Experiments with mice are currently performed at this site, using different tracers, mainly for application to oncology and to FDG physiology in the renal and liver system \cite{massollo2013metformin, garbarino2013estimate, garbarino2014novel, garbarino2015new}. In this section we describe the performance of our approach to compartmental analysis in the case of both synthetic and real 'Albira' data.

\subsection{Synthetic data}
We realized three different simulation experiments, one for each model described in this paper: a two--compartment scheme for the brain studies, a two--compartment with dual input model for the liver (coupled with a standard two--compartment scheme for the gut) and a three compartment model for renal studies. 

In order to produce the synthetic data we initially chose realistic values for V (the blood fraction that supplies the organ of interest) and for $\lambda$ of the renal system; then we utilized ground truth (g.t.) values for all the tracer kinetic parameters (4 in the brain case, 4+4 in the gut--liver case and 5 in the kidneys case). With these selected values we solved every forward problems (\ref{eq:fwdpb})--(\ref{sysmat}) in term of $C$. These solutions of the Cauchy problems are sampled in time, on a total interval $[t_1,t_m]$ that corresponds to the total acquisition time with 'Albira' and in correspondance of time points typical of experiments with the scanner in this application context. The Input Function $C_b$ has always been obtained by fitting with a gamma variate function a set
of real measurements acquired from a healthy mouse in a very controlled experiment \cite{golish2001fast}. We computed the data $\tilde{C}$ as in (\ref{eq:data}), affected the data by Poisson noise, and applied our algorithm in order to reconstruct the exchange coefficients. The Tikhonov regularization parameter $r$ was optimized at each iteration using the Generalized Cross Validation technique \cite{golub1979generalized}. Comparison with the g.t. values for this parameters provides limits about the reliability of the model and of the inversion procedure. 

The results of this test are given in Tables \ref{tab:1}, \ref{tab:2} and \ref{tab:3}, where
mean and standard deviations are computed over 50 runs of the same problem with different (random) initialization vectors $K^{(0)}$, in a Monte Carlo approach. Iterations are stopped with a discrepancy principle on data in combination with a tolerance on the step--size $H$ of the Newton Method. The computational burden is $\simeq 10$ seconds for each Monte Carlo run (on a Intel i5 2.3GHz x 4). In Tables, we show also comparison with a standard Levemberg--Marquardt method for the least--squares minimization \cite{nocedal2006numerical}. Comparison with respect to the ground--truth values and the values provided by the Levenberg-Marquardt method, and the small values for the reconstruction uncertainties clearly show the reliability of our approach.

\begin{table}
\begin{center}
\begin{tabular}{c|c|c|c|c}
\hline
BRAIN  & $k_1$ & $k_2$ & $k_3$ & $k_4$ \cr
\hline 
g. t. & 1.00  &   0.20 &   0.05 & 0.80   \cr
\hline
MGN  & $1.05\pm 0.04$   &  $0.20\pm 0.02$ & $0.05 \pm 0.01$& $0.82\pm 0.03$  \cr
\hline
LS  & $1.02\pm 0.08$   &  $0.18\pm 0.05$ & $0.05 \pm 0.05$& $0.80\pm 0.03$  \cr
\end{tabular}
\caption{Validation with a synthetic dataset: ground--truth (g.t.) values of the tracer coefficients compared with the reconstructions provided by multilinear fitting with the Levenberg-Marquardt Least-Squares algorithm (LS) and by our regularized Multivariate Gauss Newton (MGN) approach. $V$ is set equal to $0.02$. }\label{tab:1}

\end{center}
\end{table}

\begin{table}\scriptsize
\begin{center}
\begin{tabular}{c|c|c|c|c|c|c|c|c}
\hline
GUT--LIVER &$k_1^{G}$& $k_2^{G}$ & $k_3^{G}$ & $k_4^{G}(=k_v)$ & $k_a$ & $k_2$ & $k_3$ & $k_4$ \cr
\hline 
g. t.   & 1.00 &  0.20 & 0.10 & 1.00 & 1.50  &  0.40 &   0.30 &   0.60  \cr
\hline
MGN & $1.03\pm 0.03$ & $0.19\pm 0.2$ & $0.10\pm 0.01$ & $0.99\pm0.02$     & $1.52\pm 0.04$    & $0.41\pm 0.03$  &  $0.29 \pm 0.02$ &   $0.59\pm 0.01$ \cr
\hline
LS & $1.01\pm 0.07$ & $0.23\pm 0.4$ & $0.10\pm 0.07$ & $1.00\pm0.10$     & $1.51\pm 0.12$    & $0.39\pm 0.06$  &  $0.29 \pm 0.04$ &   $0.60\pm 0.07$ \cr
\end{tabular}

\caption{Validation with a synthetic dataset: ground--truth (g.t.) values of the tracer coefficients compared with the reconstructions provided by multilinear fitting with the Levenberg-Marquardt Least-Squares algorithm (LS) and by our regularized Multivariate Gauss Newton (MGN) approach. Simulated values of tracer coefficients, and reconstructed values for both the gut ($k_i^G$) and the liver coefficients. $V$ for the liver is set equal to $0.3$; $V$ for the gut is set equal to $0.1$.}\label{tab:2}
\end{center}

\end{table}

\begin{table}
\begin{center}
\begin{tabular}{c|c|c|c|c|c}
\hline
KIDNEYS &$k_1$& $k_2$ & $k_3$ & $k_4$ & $k_6$ \cr
\hline 
g. t.   & 0.80 &  0.10 & 0.20 & 1.00 & 0.70   \cr
\hline
MGN & $0.81\pm 0.03$ & $0.11\pm 0.01$ & $0.20\pm 0.03$ & $1.02\pm0.02$ & $0.72 \pm 0.03$    \cr
\hline
LS & $0.80\pm 0.10$ & $0.09\pm 0.03$ & $0.21\pm 0.05$ & $1.00\pm0.11$ & $0.70 \pm 0.09$    \cr

\end{tabular}

\caption{Validation with a synthetic dataset: ground--truth (g.t.) values of the tracer coefficients compared with the reconstructions provided by multilinear fitting with the Levenberg-Marquardt Least-Squares algorithm (LS) and by our regularized Multivariate Gauss Newton (MGN) approach. $V$ is set equal to $0.3$, and the proportional factor $\lambda$ is $10^2$ (such that $k_7 = \frac{1}{\lambda} k_4$).}\label{tab:3}
\end{center}

\end{table}

\subsection{Real FDG--PET data}

For our experiment we consider a control group (n=10) and a group (of the same size) in which FDG injection was performed after one month of high dose metformin treatment (750 mg/Kg body weight daily). This drug reduces blood glucose concentration without causing hypoglycemia \cite{klepser1997metformin}, mostly by decreasing intestinal glucose absorption and glucose delivery by the liver \cite{massollo2013metformin}. It follows that nuclear medicine experiments with such kind of animal models are perfect candidates to validate a compartmental model that has been designed in order to follow the FDG kinetics in the hepatic and renal system in a refined fashion \cite{garbarino2014novel, garbarino2015new}.

\begin{figure}
\begin{center}
\begin{tabular}{cc}
\includegraphics[width=8.5cm]{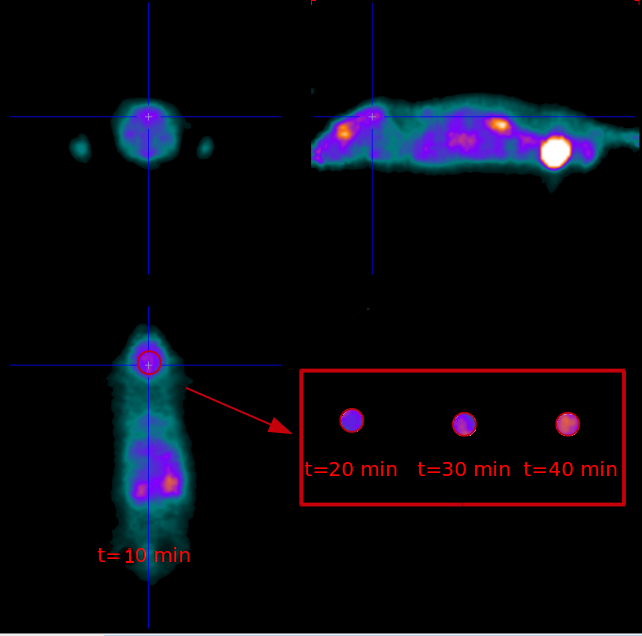}
\includegraphics[width=9.4cm]{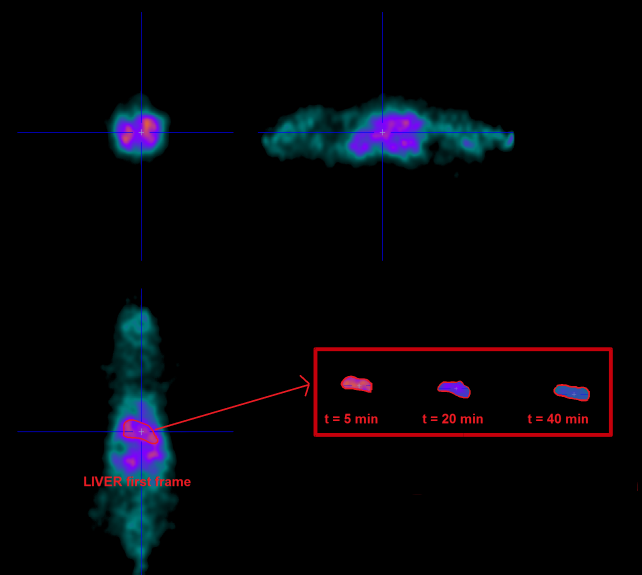}\\
\includegraphics[width=8.5cm]{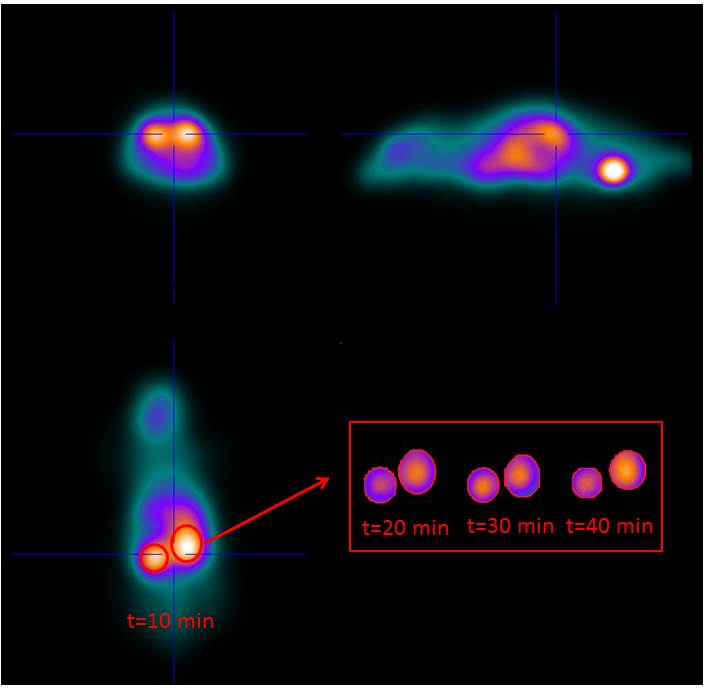}

\end{tabular}
\caption{Pictorial examples of different ROIs on different time 
points. Top-left ROIs over brain; top-right 
ROIs over liver; bottom-left ROIs over kidneys.}\label{fig:panel}
\end{center}
\end{figure}

To ensure a steady state of substrate and hormones governing glucose metabolism, the small animal was studied after six hours fasting. Mouse was weighted and anaesthesia was induced by intra-peritoneal administration of ketamine/xylazine (100 and 10 mg/kg, respectively). Serum glucose level was tested and animals were positioned on the bed of the scanner, whose two--ring configuration permits to cover the whole animal body in a single bed position. The mouse was injected with a dose of 3--4 MBq of FDG through a tail vein soon after the start of a list mode acquisition lasting 50 minutes. Acquisition was performed using the following framing rate: 10 $\times$ 15 sec, 5 $\times$ 30 sec, 2 $\times$ 150 sec, 6 $\times$ 300 sec, 1 $\times$ 600 sec. PET data were reconstructed using a Maximum Likelihood Expectation Maximization method (MLEM) \cite{hudson1994accelerated}. Thereafter, each image dataset was reviewed by an experienced observer who recognized five Regions Of Interest (ROIs) encompassing left ventricle, brain, gut and liver and kidneys respectively (as in Figure \ref{fig:panel}). The ROIs over the left ventricle allowed us to compute the IF (we are aware that the determination of IF is a challenging task in the case of mice. To accomplish it, for each animal model we have first viewed the tracer first pass in cine mode; then, in a frame where the left ventricle was particularly visible, we have drawn a ROI in the aortic arc and maintained it for all time points). The other ROIs allowed us to estimate the input data for the compartmental method, i.e. estimates of $\tilde{C}(t)$ all experimental time points. As far as $V$ is concerned, we utilized physiological sound values for each ROI, accordingly to \cite{keiding2012bringing, marzola2003vivo}. The output of our algorithm for model reduction is described in Table \ref{tab:real1}, \ref{tab:real2} and \ref{tab:real3}, which contains the mean values and standard deviations of the reconstructed tracer coefficients after 50 runs of the code for 50 random initialization vectors. 
In the hepatic case, only results for liver are shown, the one of actual physiological interest.
 
These results are physiologically plausible: for example, in all the three experiments, values of $k_3$ decrease in metformin--treated mice with respect to the control mice (the brain case is the less notable one), coherently with the property of metformin to decrease de--phosphorilation of FDG (and glucose). The relative behaviour of the coefficient values in the kidneys and liver case are coherent with the results of analogous experiments \cite{garbarino2014novel, garbarino2015new}.

\begin{table}
\begin{center}
\begin{tabular}{c|c|c|c|c}
\hline
BRAIN  & $k_1$ & $k_2$ & $k_3$ & $k_4$ \cr
\hline 
control & $0.982\pm0.063$  & $0.375\pm 0.051$ &$0.010 \pm 0.005$ & $1.082\pm0.093$ \cr
\hline
metformin  & $1.051\pm 0.916$   &  $0.264\pm 0.031$ & $0.004\pm 0.003$& $0.820\pm 0.032$  \cr
\end{tabular}
\caption{Mean and standard deviations for liver kinetic parameters over the two sets of mice (control, metformin).}\label{tab:real1}

\end{center}
\end{table}
\begin{table}
\begin{center}
\begin{tabular}{c|c|c|c|c|c}
\hline
LIVER & $k_v$ & $k_a$ & $k_2$ & $k_3$ & $k_4$ \cr
\hline 
control   &$1.620\pm 0.131$  & $1.812\pm 0.151$ & $0.003\pm 0.003$  & $0.113 \pm 0.048$ & $1.873 \pm 0.201$ \cr
\hline
metformin & $1.711 \pm 0.166$ & $1.795 \pm 0.150$ & $0.004\pm 0.003$ & $0.023 \pm 0.011$ & $1.906\pm 0.182$ \cr
\end{tabular}
\caption{Mean and standard deviations for liver kinetic parameters over the two sets of mice (control, metformin).}\label{tab:real2}
\end{center}

\end{table}

\begin{table}
\begin{center}
\begin{tabular}{c|c|c|c|c|c}
\hline
KIDNEYS &$k_1$& $k_2$ & $k_3$ & $k_4$ & $k_6$ \cr
\hline 
control   & $1.027\pm 0.109$ &  $0.193\pm 0.027$ & $0.187\pm 0.021$ & $0.415\pm 0.76$ & $0.512\pm 0.043$   \cr
\hline
metformin & $1.154 \pm 0.110$ & $173\pm 0.030$ & $0.003\pm 0.003$ & $0.490\pm 0.059$ & $0.497\pm 0.050$    
\end{tabular}

\caption{Mean and standard deviations for liver kinetic parameters over the two sets of mice (control, metformin). $k_7 = 0.042\pm 0.017$}\label{tab:real3}
\end{center}

\end{table}

\section{Conclusions}
This paper, as a sequel of the previous work concerning the compartmental inverse problem, discusses the mathematical model for the compartmental analysis and provides a general technique for its resolution. 
In particular, we described the performance of a regularized Newton approach to the numerical solution of the inverse problem associated to a compartmental model for FDG metabolism in micro--PET experiments. The numerical method is able to determine with a notable level of robustness and reliability all the tracer coefficients embedded in the model and, in particular, it does not a priori forbid de--phosphorylation. From a mathematical perspective, a further, more fundamental advantage of our scheme with respect to the Levenberg--Marquardt approach, is that Newton methods search for zeroes of non--linear functionals, and therefore do not need to a priori select a topology in the data space, as in the case of least--squares approaches. Results on real data verify the reliability of the method in estimating physiologically sound values for the tracer coefficients, that are in agreement with literature.

A Matlab prototype implementing this approach is at disposal at together with a Graphical User Interface for user-friendly input/output processing.

\section*{References}
\bibliography{biblio}

\providecommand{\newblock}{}
\begin{thebibliography}{10}
\expandafter\ifx\csname url\endcsname\relax
  \def\url#1{{\tt #1}}\fi
\expandafter\ifx\csname urlprefix\endcsname\relax\def\urlprefix{URL }\fi
\providecommand{\eprint}[2][]{\url{#2}}

\bibitem{ollinger1997positron}
Ollinger J~M and Fessler J~A 1997 {\em IEEE Signal Processing Magazine\/} {\bf
  14} 43--55

\bibitem{antoch2004accuracy}
Antoch G, Saoudi N, Kuehl H, Dahmen G, Mueller S~P, Beyer T, Bockisch A,
  Debatin J~F and Freudenberg L~S 2004 {\em Journal of Clinical Oncology\/}
  {\bf 22} 4357--4368

\bibitem{avril2001glucose}
Avril N, Menzel M, Dose J, Schelling M, Weber W, J{\"a}nicke F, Nathrath W and
  Schwaiger M 2001 {\em Journal of Nuclear Medicine\/} {\bf 42} 9--16

\bibitem{ziegler1999reproducibility}
Ziegler W~A~W~S~I, Thodtmann R, Hanauske A~R and Schwaiger M 1999 {\em J Nucl
  Med\/} {\bf 40} 1771--1777

\bibitem{delbeke2006procedure}
Delbeke D, Coleman R~E, Guiberteau M~J, Brown M~L, Royal H~D, Siegel B~A,
  Townsend D~W, Berland L~L, Parker J~A, Hubner K {\em et~al.\/} 2006 {\em
  Journal of Nuclear Medicine\/} {\bf 47} 885--895

\bibitem{engl1996regularization}
Engl H~W, Hanke M and Neubauer A 1996 {\em Regularization of inverse
  problems\/} vol 375 (Springer Science \& Business Media)

\bibitem{garbarino2013estimate}
Garbarino S, Caviglia G, Brignone M, Massollo M, Sambuceti G and Piana M 2013
  {\em Computational and mathematical methods in medicine\/} {\bf 2013}

\bibitem{garbarino2014novel}
Garbarino S, Caviglia G, Sambuceti G, Benvenuto F and Piana M 2014 {\em Physics
  in medicine and biology\/} {\bf 59} 2469

\bibitem{gunn2001positron}
Gunn R~N, Gunn S~R and Cunningham V~J 2001 {\em Journal of Cerebral Blood Flow
  \& Metabolism\/} {\bf 21} 635--652

\bibitem{kamasak2005direct}
Kamasak M~E, Bouman C~A, Morris E~D and Sauer K 2005 {\em Medical Imaging, IEEE
  Transactions on\/} {\bf 24} 636--650

\bibitem{qiao2007kidney}
Qiao H, Bai J, Chen Y and Tian J 2007 {\em International journal of biomedical
  imaging\/} {\bf 2007}

\bibitem{schmidt2002kinetic}
Schmidt K and Turkheimer F 2002 {\em The Quarterly Journal of Nuclear Medicine
  and Molecular Imaging\/} {\bf 46} 70

\bibitem{sourbron2011tracer}
Sourbron S and Buckley D~L 2011 {\em Physics in medicine and biology\/} {\bf
  57} R1

\bibitem{delbary}
Delbary F, Garbarino S and Vivaldi V {\em Inverse problems, submitted\/}

\bibitem{alf2013quantification}
Alf M~F, Wyss M~T, Buck A, Weber B, Schibli R and Kr{\"a}mer S~D 2013 {\em
  Journal of Nuclear Medicine\/} {\bf 54} 132--138

\bibitem{karakatsanis2013dynamic}
Karakatsanis N~A, Lodge M~A, Zhou Y, Wahl R~L and Rahmim A 2013 {\em Physics in
  medicine and biology\/} {\bf 58} 7419

\bibitem{wardak2013simplified}
Wardak M, Schiepers C, Wong K~P and Huang S~C 2013 Simplified reference tissue
  models for kinetic analysis of dynamic flt pet imaging in patients undergoing
  brain tumor treatment {\em Society of Nuclear Medicine Annual Meeting
  Abstracts\/} vol~54 (Soc Nuclear Med) p 1415

\bibitem{angelis2014evaluation}
Angelis G~I, Matthews J~C, Kotasidis F~A, Markiewicz P~J, Lionheart W~R and
  Reader A~J 2014 {\em Annals of nuclear medicine\/} {\bf 28} 860--873

\bibitem{garbarino2015new}
Garbarino S, Vivaldi V, Delbary F, Caviglia G, Piana M, Marini C, Capitanio S,
  Calamia I, Buschiazzo A and Sambuceti G 2015 {\em EJNMMI Research\/} {\bf 5}
  1

\bibitem{delbeke1999optimal}
Delbeke D, Rose D~M, Chapman W~C, Pinson C~W, Wright J~K, Shyr R~D~B~Y and
  Leach S~D 1999 {\em J Nucl Med\/} {\bf 40} 1784--1791

\bibitem{diederichs1998fdg}
Diederichs C~G, Staib L, Glatting G, Beger H~G and Reske S~N 1998 {\em The
  Journal of Nuclear Medicine\/} {\bf 39} 1030

\bibitem{shreve1999pitfalls}
Shreve P~D, Anzai Y and Wahl R~L 1999 {\em Radiographics\/} {\bf 19} 61--77

\bibitem{massollo2013metformin}
Massollo M, Marini C, Brignone M, Emionite L, Salani B, Riondato M, Capitanio
  S, Fiz F, Democrito A, Amaro A {\em et~al.\/} 2013 {\em Journal of Nuclear
  Medicine\/} {\bf 54} 259--266

\bibitem{golish2001fast}
Golish S~R, Hove J~D, Schelbert H~R and Gambhir S~S 2001 {\em Journal of
  Nuclear Medicine\/} {\bf 42} 924--931

\bibitem{golub1979generalized}
Golub G~H, Heath M and Wahba G 1979 {\em Technometrics\/} {\bf 21} 215--223

\bibitem{nocedal2006numerical}
Nocedal J and Wright S 2006 {\em Numerical optimization\/} (Springer Science \&
  Business Media)

\bibitem{klepser1997metformin}
Klepser T and Kelly M 1997 {\em American journal of health-system pharmacy\/}
  {\bf 54} 893--903

\bibitem{hudson1994accelerated}
Hudson H~M and Larkin R~S 1994 {\em Medical Imaging, IEEE Transactions on\/}
  {\bf 13} 601--609

\bibitem{keiding2012bringing}
Keiding S 2012 {\em Journal of Nuclear Medicine\/} {\bf 53} 425--433

\bibitem{marzola2003vivo}
Marzola P, Farace P, Calderan L, Crescimanno C, Lunati E, Nicolato E, Benati D,
  Degrassi A, Terron A, Klapwijk J {\em et~al.\/} 2003 {\em International
  journal of cancer\/} {\bf 104} 462--468

\end{thebibliography}
\end{document}